\documentclass[11pt]{aa}

\usepackage[utf8]{inputenc}
\usepackage[T1]{fontenc}

\usepackage[varg]{txfonts}

\usepackage{float, graphicx}
\usepackage{placeins}

\usepackage{multirow}

\usepackage{subfigure}
\usepackage{lipsum}

\usepackage{color}
\usepackage{ulem}

\bibpunct{(}{)}{;}{a}{}{,} 

\title{K-Stacker: Keplerian image recombination for the \\direct detection of exoplanets}
\titlerunning{Keplerian image recombination for the direct detection of exoplanets}

\author{M. \text{Nowak}\inst{\ref{lesia}, \ref{lam}}
  \and H. \text{Le Coroller}\inst{\ref{lam}}
  \and L. \text{Arnold}\inst{\ref{ohp}}
  \and K. \text{Dohlen}\inst{\ref{lam}}
  \and D. \text{Estevez}\inst{\ref{gre}, \ref{lam}}
  \and T. \text{Fusco}\inst{\ref{onera}}
  \and J.-F. \text{Sauvage}\inst{\ref{onera}}
  \and A. \text{Vigan}\inst{\ref{lam}}}
  
\authorrunning{M. Nowak, H. Le Coroller, et al.}
\titlerunning{Keplerian image recombination for the direct detection of exoplanets}

\institute{LESIA, Observatoire de Paris, Université PSL, CNRS, Sorbonne Université, Univ. Paris Diderot, Sorbonne Paris Cité, 5 place Jules Janssen, 92195 Meudon, France \label{lesia}\\
\email{mathias.nowak@obspm.fr}\label{isae}
\and Laboratoire d’Astrophysique de Marseille, Pôle de l’Étoile Site de Château-Gombert, 38 rue Frédéric Joliot-Curie 13388 Marseille Cedex 13, France \\    
\email{herve.lecoroller@lam.fr}\label{lam}
\and Aix Marseille Université, CNRS, OHP (Observatoire de Haute Provence), Institut Pythéas UMS 3470, 04870 Saint-Michel-l’Observatoire, France \label{ohp}
\and ONERA - The French Aerospace Lab BP72 - 29 avenue de la Division Leclerc FR-92322 Chatillon Cedex \label{onera}
\and Univ. Grenoble Alpes, Univ. Savoie Mont Blanc, CNRS, LAPP, 74000 Annecy, France \label{gre}}

\abstract{Angular Differential Imaging (ADI) takes advantage of the field rotation naturally induced by altitude-azimuth mounts to reduce static speckle noise. Used with facilities like SPHERE at the VLT, this technique allows to achieve contrast ratios of $10^{-6}$. But the ADI method intrinsically limits the useful exposure time on a given target (to about 1-2 h per night). Detecting fainter exoplanets requires to be able to combine multiple observations acquired on different nights, potentially spread on several weeks or months. But the unknown orbital motion of the planet makes it particularly diffcult to properly combine all observations. In the near future, with the upcoming generation of Extremely Large Telescopes (ELTs) with increased resolution, the orbital motion may even become a problem on a single night.}
{We present a proof of concept for a new algorithm which can be used to detect exoplanets in high contrast images. The algorithm properly combines mutliple observations acquired during different nights, taking into account the orbital motion of the planet.} 
{We simulate SPHERE/IRDIS time series of observations in which we blindly inject planets on random orbits, at random level of S/N, below the detection limit (down to $\mathrm{S/N}\simeq1.5$). We then use an optimization algorithm to ``guess'' the orbital parameters, and take into account the orbital motion to properly recombine the different images, and eventually detect the planets.}
{We show that an optimization algorithm can indeed be used to find undetected planets in temporal sequences of images, even if they are spread over orbital time scales. As expected, the typical gain in S/N ratio is $\sqrt{n}$, $n$ being the number of observations combined. We find that the K-Stacker algorithm is able de-orbit and combine the images to reach a level of performance similar to what could be expected if the planet was not moving. We find recovery rates of $\simeq{}50\%$ at S/N=5. We also find that the algorithm is able to determine the position of the planet in individual frames at one pixel precision, even despite the fact that the planet itself is below the detection limit in each frame.}
{Our simulations show that K-Stacker can be used to detect planets at very low S/N level, down to $\simeq$~2 in individual frames, for series of 10 images. This could be used to increase the contrast limit of current exoplanet imaging instruments and to discover fainter bodies. We also suggest that the ability of K-Stacker to determine the position of the planet in every image of the time serie could be used as part of a new observing strategy in which long exposures would be broken into shorter ones spread over months. This could make possible to determine the orbital parameters of a planet without requiring multiple high S/N >5 detections.}


\begin{document}

\maketitle{}

\section{Introduction}


The direct detection of exoplanetary bodies is an extremely challenging task: for a planet orbiting at 10 AU around a star at a few tens of parsec, the typical angular separation is about 0.1 to 1 arcsec, and typical contrast ratios in the near-infraRed (NIR) are expected to be of the order of $\approx 10^{-5}$ for a young Jupiter-like planet \citep{Marley2007} to $10^{-10}$ for an Earth-like planet.

The recent development of eXtreme Adaptive Optics (XAO) systems, which are now able to deliver high Strehl ratios by correcting high-order wavefront errors, as well as of coronagraphic imaging systems has lead to a new generation of high-contrast imagers (GPI, \citeauthor{Macintosh2014}, \citeyear{Macintosh2014}; SPHERE, \citeauthor{Beuzit2008}, \citeyear{Beuzit2008}). However, even with these state-of-the-art facilities, observations are still crippled by speckle noise, originating in atmospheric turbulence and instrumental defects. Since atmospheric speckles have very short decorrelation times, they can be averaged by increasing the exposure time. The situation is a little different when dealing with the so-called pseudo-static (instrumental) speckles, {which have much longer correlation times}. Without any specific strategy, static speckles typically limit the useful exposure time to a mere handful of seconds, in the most favorable cases (\citealt{Macintosh2005, Soummer2007, Hinkley2007}).

Numerous image processing techniques have been proposed to mitigate this static speckle noise, and some are currently being applied for example to SPHERE and GPI data with great success. But subtracting static speckles without affecting any potential exoplanet light is not easy. Some methods \citep{Sparks2002, Thatte2007, Racine1999} use the fact that, contrary to a planet, the distance of any given speckle from the center of the star scales as $\lambda/D$, where $\lambda$ is the wavelength of observation. This allows to disentangle the speckles from the planet, and subtract them. On a similar line of thought, one can also use the difference of polarization between the unpolarized starlight and the polarized planet signal to distinguish them, as it is done with the Differential Polarimetry method \citep{Canovas2011}. Finally, requiring absolutely no spectral or polarimetric capability whatsoever, the Angular Differential Imaging (ADI, \citealt{Marois2006}) uses the rotation of the Field of View (FoV) naturally induced by alt-az mounts to disentangle pseudo-static speckles from any moving planet.

The ADI and SDI methods are currently implemented in the SPHERE data reduction pipeline, allowing to routinely achieve contrast ratios of $10^{-5}$ and up to $10^{-6}$ in excellent conditions \citep{Zurlo2014, Vigan2015, Zurlo2016}.
However, when using this ADI method, the total time for an observation is limited by the FoV rotation to $\simeq{}1\textrm{h}$, a limitation wich was already pointed out by \citet{Marois2006}.

An easy way to improve the detection limit of any astronomical instrument is usually to use longer exposure times. But when using a technique like ADI, this can rapidly turn to a very intricate problem: if exposures are limited to about an hour per night, gathering for example 10 to 20 hours of exposure require observing on 10 to 20 different nights. Because of weather and observatory constraints, this can rapidly lead to extended time sequences, where the different images are possibly taken over different observing runs, with several month intervals. In such a case, the orbital motion of the planet, which up until now has almost always be ignored when combining images, must be taken into account. This is especially true when using large telescope, probing the innermost part of close-by stellar system. In such a case, even data acquired during succeeding nights {can be affected by the orbital motion of the planet (see \citeauthor{Males2013}, \citeyear{Males2013} for a detailed disccussion of this particular problem)}

{\cite{Males2015} also worked on the problem of recombining images acquired on longer timescale, in which the planet was moving on a significant part of its orbit. However, they use the assumption that the planet, albeit very faint, could be seen in each individual frame. In this regime, deorbiting can help improve the S/N of the planet, but cannot help detecting new planets, which could not be seen in a single frame. Males et al. themselves stated that ``an important area of investigation will be the performance of [ODI] when there is no prior information with which to determine the orbits.''}

{This is precisely the problem we intend to tackle with the K-Stacker algorithm.}

In a previous paper \citep{2015arXiv151006331L}, we have proposed this new technique to detect exoplanets, called K-Stacker, which could allow to combine multiple observations made over different nights, to increase the contrast limit of direct imaging instruments. 
In the present work we describe in detail the K-Stacker algorithm and we provide a statistical analysis intended as a ``proof of concept'' of this method.



In Section~\ref{sec:simulations}, we describe the simulated images on which we tested our K-Stacker algorithm. Section~\ref{sec:algorithm} focuses on the problem of recombining multiple observations in which the planet remains undetected, and gives a description of our algorithm, 
based on a combination of a brute-force and a gradient-descent methods to determine the orbital parameters. In Section~\ref{sec:results}, we present the results of a blind test performed on 50 independant and random simulated observation with this algorithm.
We discuss the performances of the algorithm and the required computer resources in Section~\ref{sec:discussion}. Our conclusions are given in Section~\ref{sec:conclusion}.

\section{Simulated data}
\label{sec:simulations}

In order to develop and test the K-Stacker algorithm, we developed a simple simulator to generate similar close to the one produced by the SPHERE/IRDIS instrument \citep{Dohlen2008} in its dual-band imaging mode \citep{Vigan2010}, in terms of general characteristics (size of the FoV, width of the PSF, etc.). Whereas we believe that K-Stacker could benefit from an ADI/SDI pre-processing step which remove most of the pseudo-static speckles, we do not simulate ADI or SDI reduced data.

Even if a number of studies \citep{Hinkley2007, Soummer2007, Martinez2012, Martinez2013} have been done on the temporal evolution of instrumental speckles, the behaviour of these quasi-static speckles is hard to simulate because they are related to many different factors  (thermal conditions, mechanical stress, instrumental configuration, etc.).

In the present work, we focuse on the case of pure atmospheric speckles behind an XAO, which are completely uncorrelated from one image to the other. We also assume that the coronagraph is perfect, and that the coherent part of the light is fully removed. If we do not take into account variations in amplitude of the incoming wavefront, a perfect coherent light suppression can be defined analytically using the following equation \citep{Fusco2006}:
 
\begin{equation}
I(x, y)=\left|\mathrm{F}\left[\left(e^{i\Delta{}\phi(a, b)}-e^{\frac{1}{2}{\sigma_{\Delta{}\phi}}^2}\right)P(a, b)\right](x, y)\right|^2
\label{eq:coronagraph}
\end{equation}
\noindent{}where $I(x, y)$ represents the intensity in the focal plane, $P(a, b)$ the aperture function, $\Delta{}\phi(a, b)$ the phase error of the wavefront, F the Fourier transform, and $\sigma_{\Delta{}\phi}$ the standard deviation of the phase error in the incoming wavefront.

In the above equation, the term $\Delta{}\Phi$ is what is ultimately responsible for the speckle noise in the final images, which is known to originate in atmospheric phase variations and/or instrumental defects.


The atmospheric phase masks were simulated using an IDL code developed by \cite{Fusco2006}, which generates phase errors downstream of the adaptive optics system. 

The images generated are $512\times{}512$ pixels, and only represent a part of the bigger SPHERE detector. Our code generates monochromatic images, with a $12.25\text{ mas/pixel}$ spatial sampling corresponding to SPHERE/IRDIS.

\begin{figure}
\begin{center}
\includegraphics[width=\linewidth]{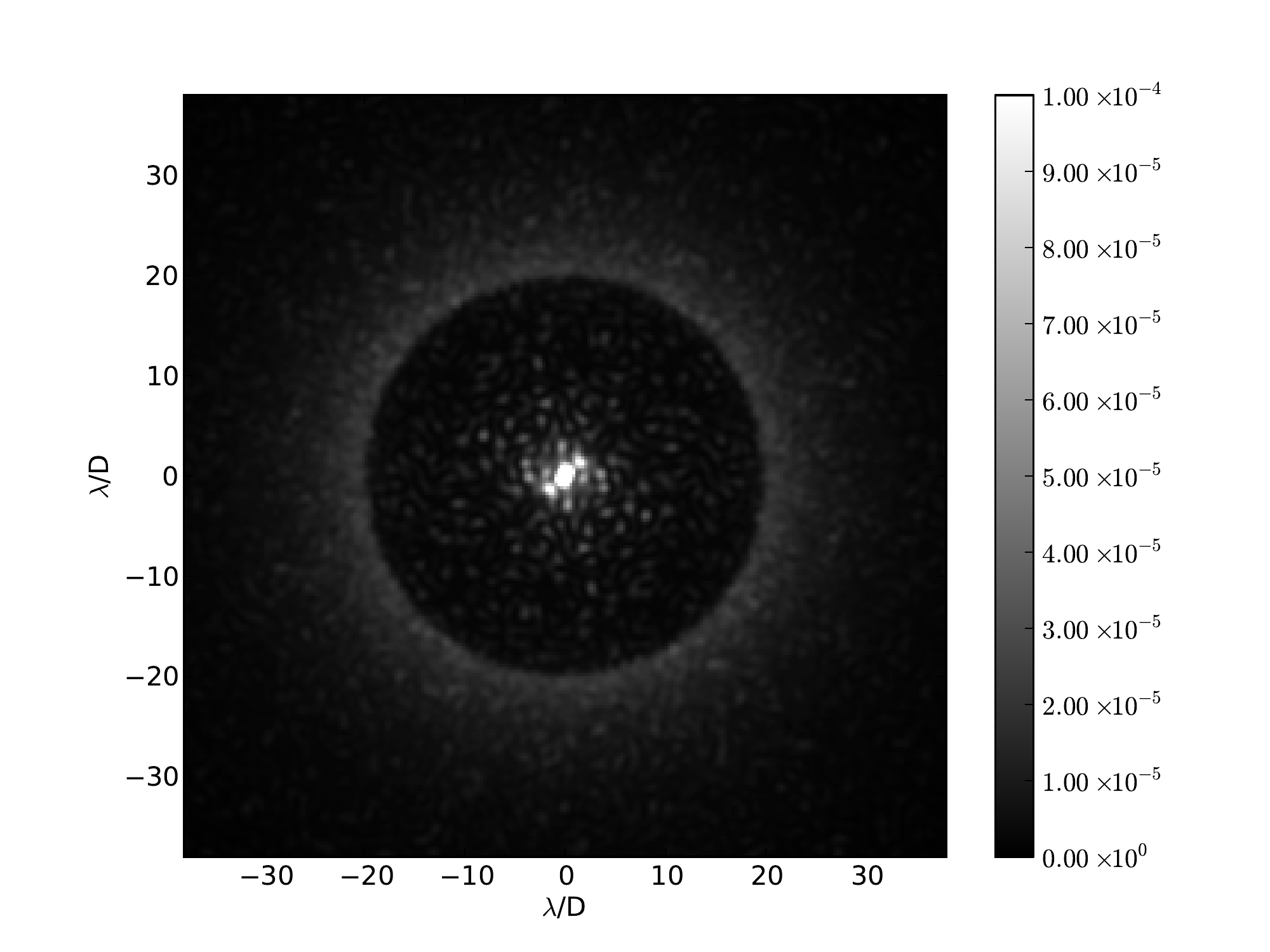}
\caption{Simulation of an AO corrected image at $1.6~\mu{}\mathrm{m}$ for SPHERE/IRDIS, under a 0.8'' seeing sky. The AO corrected area is clearly visible, and extends to about $20\lambda/D$ (1.6''). The image is scaled to the peak value of the non-coronographic star PSF.}
\label{fig:noisy_alc}
\end{center}
\end{figure}

In Figure~\ref{fig:noisy_alc}, {we show one of our typical exposure, made with our simulated SPHERE/IRDIS instrument, under a 0.8'' seeing sky. This image has been obtained by averaging 100 exposures made with 100 uncorrelated atmospheric masks}. The image is normalized to the central peak intensity of the non-coronagraphic PSF. As it was unecessary for our work, we did not try to calibrate our simulations to match the correct SPHERE/IRDIS photometry, and thus {did not include photon noise}.

Finally, false planets are injected by directly adding a non-coronagraphic PSF in the images. This means that neither the coronagraph nor the atmospheric and/or instrumental phase errors have any impact on the exoplanet signal.

\section{Recombining the images: the K-Stacker algorithm}
\label{sec:algorithm}

\subsection{K-Stacker principle: the recombination as an optimization problem}

Consider a set of $n$ images $I_1, \dots{}, I_n$ of a given star, acquired at times $t_1, \dots{}, t_n$, and suppose that an exoplanet is orbiting around this star. In the $k$-th image, the planet is at position $(x_k, y_k)$, but remains undetected. Thanks to the law of orbital mechanics, these positions $x_k, y_k$ can be related to 7 parameters: the 6 orbital elements, i.e. eccentricity $e$, semi major-axis $a$, epoch at perihelion $t_0$, longitude of the ascending node $\Omega$, inclination $i$, argument at periapsis $\theta_0$, and the mass of the central star $M_*$. The distance of the star $d_*$ is also required to project the orbit on the CCD.

K-Stacker is based on the idea that when trying to recombine the images to detect a hidden orbiting planet, a strong spot should emerge only when the images are recombined along the correct orbit. Otherwise, the speckles should just average to a certain value, within statistical fluctuations.

To recombine the images along a given orbit, we first compute the position of the planet predicted by the laws of orbital mechanics at times $t_1$ in image $I_1$, at time $t_2$ in image $I_2$, etc. Then, we rotate/translate each image $I_k$ for $k>0$ to align all these positions, and add all the images. This means that when we recombine the images, we actually know where we expect to see a strong feature if the orbital parameters are correct, which suggests the use of a Signal-to-Noise Ratio (S/N ratio) at this particular position as a measure of the ``quality of the recombination''.


For a given orbit, the position of a planet at a time t is defined by $x=(t, a, e, t_0, \Omega, i, \theta_0, M_*)$
and assuming that the noise in the different images are fully uncorrelated, the S/N ratio can be computed in the following way: 
\begin{itemize}
  \item{First, for each image $I_k$, we compute the position $(x_k, y_k)$ by solving the Kepler equation with a Newton algorithm, projecting the orbital position on the detector plane. This gives a set of one position per image, which can be expressed in polar coordinates: $(r_1, \theta_1), \dots{}, (r_n, \theta_n)$.}
  \item{Then, for each $k$ in $\{1, \dots, n\}$, we compute the flux at position $(r_k, \theta_k)$ in the image $I_k$, using a circular integration box, of radius equals to the Full Width at Half-Maximum (FWHM) of the non-coronagraphic PSF of the instrument. To this value, we subtract the estimate of the background flux in image $I_k$ at radius $r_k$. 
  This gives the signal in image $k$, $s_k(x)$ at the position $(r_k, \theta_k)$. The total signal is the sum of all of these values in each image $S(x)=\sum_k{s_k(x)}$.}
  \item{Finally, for each $k$ in $\{1, \dots{}, n\}$, we compute the noise level $\sigma_k$ at radius $r_k$ in image $I_k$, and add these values quadratically to obtain the global noise value : $N(x)=\sqrt{\sum_k{\sigma_k}^2}$. The signal to noise ratio equals: $\mathrm{S/N}(x)=S(x)/N(x)$.}
\end{itemize}

This S/N, seen as a function of the 6 orbital parameters (and of the star mass $M$ and distance $d$, if unknown), can be optimized using a modified brute-force algorithm to detect planets hidden in the individual frames $I_k$.

\subsection{Modified brute-force optimization}
\label{sec:gradient}
\subsubsection{Brute force stage}


We use a brute-force algorithm to search for the maximum $S/N$.
We keep a simple shape for the search grid (6-dimensional rectangle, with linear sampling). The algorithm may therefore lose some time exploring unrealistic possibilities.
We explore the entire range of possible values, from $-\pi$ to $+\pi$ for $\Omega$ and $\theta_0$, and from 0 to $\pi$ for $i$. For $e$, we explore values ranging from $0$ to $0.8$. For $t_0$, we also explore the entire possible range, i.e. from $0$ to $T$, $T$ being the orbital period. Since the value of the orbital period depends on $a$, we take the largest interval: from $0$ to $2\pi{}\times{}\sqrt{\frac{{a_{max}}^3}{GM}}$. For $a$, we chose to explore values ranging from \hbox{$0.09\times{}d$} to $0.75\times{}d$, which correspond to the A.O. corrected area of SPHERE/IRDIS, $d$ being the distance of the star in parsecs.

The smaller the step size of the grid is, the higher chances are to find the true maximum, but the longer the computation time is. We estimated the best sampling for each parameter empirically, by looking at the typical width of the global maximum in different configurations. 


In Table~\ref{tab:grid}, we summarize the different characteristics of our grid, built for analysing observations of a star of mass $1~M_{\odot}$, located at 10~pc. Whereas the step sizes and number of points to use are only rough estimates, this table shows that the total number of points to be explored is of the order of $10^8$. The simulations have shown that such a grid allows to find a solution in a reasonable computation time (see Sect. \ref{sec:discussion}).

\begin{table}
\begin{tabular}{c c c c c c}
\hline
\hline
Param. & Unit & Min val. & Max val. & Step size & Points\\
\hline
$a$ & AU & 0.9 & 7.5 & 0.7 & 10\\
$e$ & - & 0 & 0.8 & 0.08 & 10\\
$t_0$ & yr & 0 & 20 & 0.2 & 100\\
$\Omega$ & rad & $-\pi$ & $+\pi$ & 0.18 & 35\\
$i$ & rad & $-\pi$ & 0 & 0.5 & 7\\
$\theta_0$ & rad & $-\pi$ & $+\pi$ & 0.18 & 35\\
\hline
\end{tabular}
\caption{Main characteristics of the grid used by our brute-force algorithm, computed for a star of mass $M_\odot{}$ at 10 pc.}
\label{tab:grid}
\end{table}

\subsubsection{Gradient-descent re-optimization stage} 

The weak point of the brute-force algorithm is that it may miss the global maximum, likely to be between the points of the finite grid sampling.
To circumvent this particular issue, we add a gradient-descent optimization stage to the process, and re-optimize some of the best solutions found by the brute-force algorithm. To decide how many of the best grid values should be re-optimized, we ran the brute-force algorithm on a set of 25 images generated using our SPHERE/IRDIS simulator (see Sect. \ref{sec:simulations}), in typical sky conditions (seeing of 0.8''), in which no planet was introduced. This gave us a typical distribution of the noise values sampled by the grid, in which we found that the {highest S/N value was 5.2, and that only 100 S/N values are higher than 4.5 and 1000 higher than 4.2}. This means that
if we re-optimize the best value found on the grid, a planet will be detected only if the corresponding S/N grid-maximum reaches a value higher than 5.4. However,
if we re-optimize the 100 (resp. 1000) best values, then the planet will be detected if the corresponding grid value is greater than 5.1 (resp. 4.7). Compared to the total $10^8$ points of the grid, re-optimizing 100 or 1000 values takes only little time, and allow to recover low S/N planets.


In conclusion of this section, we built an algorithm which works in three steps:
\begin{itemize}
\item{First, a brute-force algorithm is used to determine the value of the S/N function in each point of the grid.}
\item{Then, the $p=100$ highest values found are re-optimized by a gradient-descent algorithm or similar. We use the Broyden-Fletcher-Goldfarb-Shanno method, \citeauthor{BFGS} \citeyear{BFGS}.}
\item{Finally, we search for planets in all images corresponding to the re-optimized S/N values above 4.7.}
\end{itemize}

\section{Results}
\label{sec:results}

\subsection{Blind test}

To test the K-Stacker algorithm, we performed a blind experiment. We used the SPHERE/IRDIS simulator presented in Section~\ref{sec:simulations} to build 50 sets of 10 independent simulated observations, with a seeing value of 0.8''. In each set, a planet was then injected on a random orbit, at a random S/N level. The orbit was selected by drawing the 6 parameters randomly using the distributions given in Table~\ref{tab:parameters} allowing the algorithm to redraw a new set of parameters when the first one resulted in an orbit going outside of the AO corrected area in at least one image.{The mass and the distance of the star are expected to be known with a good accuracy, as it is usually the case for the bright stars observed in coronagraphy.} {For each simulation, the planet was not injected at ``constant S/N'', but rather at ``constant flux'', to simulate what is expected from a constant exposure time serie of observations. For each serie of 10 observations, the noise level $\sigma_1$ at the position of the planet in the first image is computed, and the planet is injected at a flux value of $F = \mathrm{snr}_1 \times{} \sigma_1$ on top of the background in each image, where $\mathrm{snr}_1$ is randomly selected among four possible values : 0, $5/\sqrt{10}$, $7.5/\sqrt{10}$, or $10/\sqrt{10}$.}

{Because the local value of the background behind the planet varies in each observation, as does the noise level if the planet is moving with respect to the central star, the exact S/N ratio differs from one image to the other, and the total expected S/N level of the simulation can only loosely be expected to be $S/N \simeq{} \sqrt{10} \times{} \mathrm{snr}_1$}

In all these simulations the star has a mass of $1M_{\odot{}}$, is located at a distance of 10~pc, and has a magnitude of 8 in the R band (AO sensing band). 
{The ten images of each set correspond to observations made at different times, selected arbitrarily to represent a plausible sequence of K-Stacker observations. The times are given in Table~\ref{tab:times}.}

The 50 sets of observations have been prepared using a dedicated computer program, which drew all the random parameters and S/N values, and stored them in a file. All the simulations were then processed by the K-Stacker algorithm. For each simuation, K-Stacker produced a set of 100 best optimized recombinations, with their associated S/N values (before and after the re-optimization step) and orbital parameters. {The observer has absolutely no idea of the S/N and orbital parameters of the planets injected in the data.}

{The final images produced by K-Stacker can be checked by the observer, as usually done with high-contrast images. The observer can look at the final recombined images and, based on the total recombined S/N and the shape of the detection found by K-Stacker, assign a flag to each set of observation, validating or not the potential planet candidate.}

{The file containing the parameters and S/N levels was only retrieved at the very end of the whole process
in an effort to ensure that the observer never knew in advance what to expect from each simulation, and reduce overall bias.}

\begin{table}
\begin{center}  
  \begin{tabular}{c l l}
    \hline
    \hline
    Parameter & Range & Distribution\\
    \hline
    $M_{\text{star}}$ & $1_{M\odot{}}$ & fixed value \\
    $d_{\text{star}}$ & 10 pc & fixed value \\    
    $a$ & [0.2 A.U., 7.5 A.U.] & uniform \\
    $e$ & [0, 0.5]  & uniform \\
    $t_0$ & [-20 yr, 0 yr] & uniform \\
    $\Omega$ & [-180 deg, 180 deg] & uniform \\
    $i$ & [0, 180 deg] & uniform \\
    $\theta_0$ & [-180 deg, 180 deg] & uniform \\
    $\sqrt{10}\times{}S/N$ & \{0, 5.0, 7.5, 10.0\} & uniform \\        
    \hline
  \end{tabular}
  \caption{Parameters used to inject the planet in the 50 simulation of our blind experiment.}
  \label{tab:parameters}
\end{center}  
\end{table}

\begin{table}
  \begin{center}
    \begin{tabular}{c l}
      \hline
      \hline
      Observation number & Date \\
      \hline
      1 & January, 1, 2017\\
      2 & February, 6, 2017\\
      3 & March, 15, 2017\\
      4 & April, 20, 2017\\
      5 & Februray, 6, 2018\\
      6 & March, 15, 2018\\
      7 & February, 6, 2019\\    
      8 & March, 15, 2019\\
      9 & Februray, 6, 2020\\
      10 & March, 14, 2020\\
      \hline
    \end{tabular}      
      \caption{Dates used for each of the 10 observations constituting every set of the blind test}
      \label{tab:times}
  \end{center}
\end{table}

\subsection{Results obtained}

\begin{figure}
  \begin{center}
    \includegraphics[width = \linewidth]{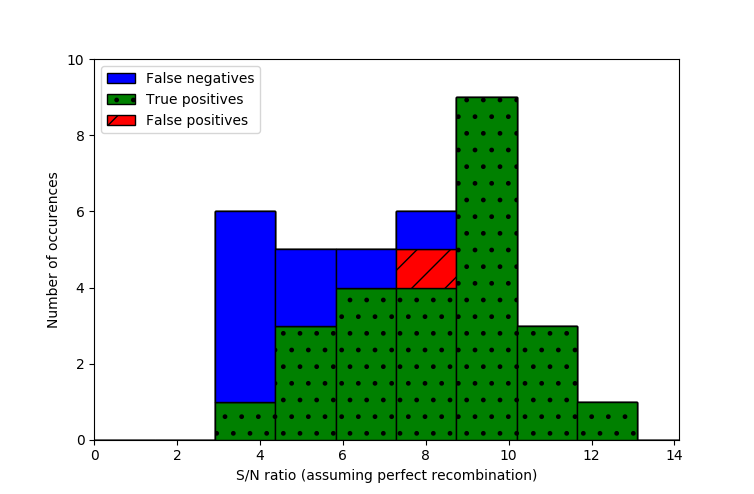}
    \caption{Distribution of the planet candidates found and missed among the 49 simulations as a function of the S/N given a perfect recombination of the images. The true negatives, all grouped at S/N = 0 are not shown.}
    \label{fig:candidates}
  \end{center}
\end{figure}

Among the 50 simulations prepared and analysed in our blind experiment, 15 were assigned by our algorithm to a group of $S/N \simeq 0$, 12 to $S/N \simeq 5$, 10 to $S/N \simeq 7.5$, and 13 to $S/N \simeq 10$.
All the 15 simulations in which no planet was actually injected were correctly identified as containing no planet candidate by the observer. In the remaining 35 simulations, 25 planets were correctly identified as planet candidates by the observer, 9 planets were missed, and one false positive was found. In Figure~\ref{fig:candidates}, we show the distribution of the planets missed/found as a function of the total real S/N ratio of the simulation computed afterwards by combining the images using the true set of orbital parameters.

\section{Discussion}
\label{sec:discussion}

\subsection{Success rate and required computer ressources}

Fig~\ref{fig:candidates} indicates that very high success rates (100\%) can be expected of K-Stacker when the number of images acquired and/or the planet signal strength is high enough so that the total recombined S/N can reach at least a value of 9. When the S/N is lower (down to 6), the algorithm can still detect the planet in most cases ($\simeq{}80\%$ in total). When the S/N lies around 5, a 50\% recovery rate seem to be achievable using the K-Stacker algorithm. This is very encouraging, as it is comparable to what can be expected using more conventional techniques.

These success rates have been obtained using the computer cluster at Laboratoire d'Astrophysique de Marseille (LAM). We used a total of 50 cores of the cluster for an estimated total computing power of about 150~GFlops, and each of the simulation (10 images) took about 10 to 15 hours for the grid search, plus about 1 to 2 hours for the gradient-descent reoptimization. These numbers scale linearly with the number of frames processed in an observation.


\subsection{The problem of false positives}
\label{sec:false_positives}

The success rate of a planet search algorithm, which defines its ability to effectively find planets and avoid ``false negatives'', is an important characteristics. But its ability to avoid ``false positives'' (FP) where a speckle is falsely flagged as a planet, is also fundamental.

In a standard single-frame high-contrast observation, the probability of getting a false detection can be written as the product of the typical number of speckles in the images and the probability of any one speckle to be luminous enough to be mistaken for a planet:
\begin{equation}
  P_\mathrm{FP}=N_\mathrm{speckles}\times{}P\left(\text{S/N speckle}>\text{S/N}_t\right)
\end{equation}
\noindent{}where $\text{S/N}_t$ is the threshold S/N ratio, corresponding to the minimum value of S/N a feature in the image has to reach in order to be considered as a potential planet.

Considering only the AO corrected area, an instrument like SPHERE/IRDIS typically has a number of speckles $N_\mathrm{speckles}\approx 1000$. Assuming that speckle noise due to atmospheric turbulence follow a normal distribution of variance 1 (in unit of S/N), the broadly used value of $\text{S/N}_t=5$ lead to a probability of false detection of $\simeq{}3\times10^{-4}$, or about one false detection every 3500 observations.

At first glance, in the case of K-Stacker, the situation seems much more problematic since each of the $10^8$ orbits tried by the algorithm can potentially lead to a false detection. This particular problem was already pointed out by \cite{Males2013}. A first estimate of the false positive probability can be done assuming that the S/N value of each orbit tried by K-Stacker is independant of all the other values.  In this case, the probability of getting a false detection on any K-Stacker run can be written as:
\begin{equation}
  \label{eq:pfd}
  P_\mathrm{FP}=1-\left[1-P\left(\text{S/N orbit}>\text{S/N}_t\right)\right]^{N_\mathrm{orbits}}
\end{equation}
If we also assume that each orbit leads to an S/N value distributed according to a normal law of mean 0 (counting the background subtraction) and variance 1 (see Sect~\ref{sec:algorithm}), then the threshold ${\text{S/N}}_t=5$ leads to $P(\text{S/N orbit}>S/N_t)=2.9\times{}10^{-7}$, and the probability of getting a false positive is almost exactly 1.\\

\begin{figure}
  \begin{center}
    \includegraphics[width = \linewidth]{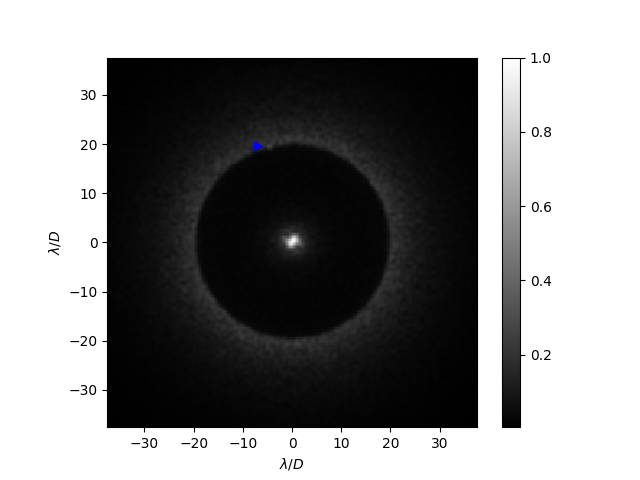}
    \caption{One of the individual frame of simulation 48, which lead to the only fase positive of the serie. The planet that has not been detected is on the edge of the AO corrected area (indicated by the blue arrow).}
    \label{fig:false_positive}
  \end{center}
\end{figure}

However, in all the 50 simulations of our blind test, only one gave a false positive result. In this simulation, a planet was injected at a high S/N value (about 2 to 3 in each individual frame), but very close to the edge of the AO corrected area (see Fig~\ref{fig:false_positive})

\begin{figure}
  \begin{center}
    \includegraphics[width = \linewidth]{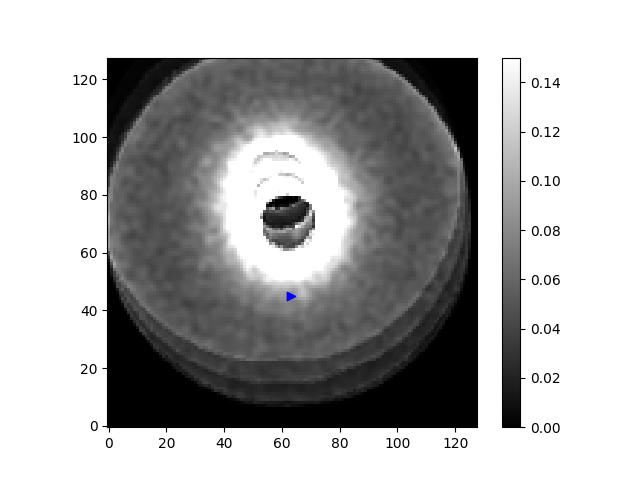}
    \caption{Recombined K-Stacker image where the false planet candidate can be seen (indicated by the arrow). The observer noted the assymetry of the planet spot.}
    \label{fig:false_positive_recombination}    
  \end{center}
\end{figure}

The planet itself was not detected, but another feature was falsely flagged as a planet candidate. However, the observer also noted that S/N value found by K-Stacker was low (4.80 before the reoptimization, 4.98 after), and that the shape of the recombined spot had an apparent lack of central symmetry (Fig~\ref{fig:false_positive_recombination}). The result was flagged as a planet candidate, but with a comment saying that it should be taken with caution. There is no doubt that if this case was a real one, the observer would have requested further observations before claiming a planet detection.

{At this point, it must also be pointed out that with recent instruments like SPHERE, which provide images at several wavelengths (IFS), color-magnitude diagrams are usually used to help discriminate between potential planet candidates and brown dwarves, background features, or speckles. In our monochromatic blind test, this check cannot be done, and the rate of false detections is necessarily higher than in classical coronagraphy, which makes our result pessimistic.}

Overall, the K-Stacker algorithm seems to be much more resilient to false positives than what could be expected from Eq. \ref{eq:pfd}. We believe that two reasons can explain this.

Firstly, the previous reasoning assumes that all the orbits tested by K-Stacker are independant. This leads to a very high ``number of trials'', and thus to a high false positive probability. It is unlikely, though, that all the $10^8$ orbits tested could really be independant. Different sets of orbital parameters can correspond to very similar orbits, especially when the total time spanned by the observations is small compared to the orbital period. This effectively reduces the number of independant orbits that should be taken into account in Equation~\ref{eq:pfd}. {We did not try to thouroughly test this hypothesis, but interestingly enough, this idea, based only on the empirical results of our K-Stacker runs and on our experience using it, agrees with the conclusion reached by \cite{Males2013}, using a more sophisticated theoretical reasoning.}

Secondly, we also noticed that when the optimization algorithm ends on a ``noise maximum'' which could lead to a false positive, the resulting image does not show a clear PSF-shaped spot, as it is the case when the algorithm finds the planet maximum (see Appendix~\ref{app:solutions} for some examples). In Table~\ref{tab:no_planet_snr}, we give, for each of the 15 simulations in which no planet was injected, the highest S/N value of the 100 best solutions found during the grid search as well as the minimum and maximum S/N values found by the gradient-descent algorithm. This table shows that if a simple S/N threshold were to be used to flag planet candidates, the number of false positives would be much higher. This result emphasizes the possibility of using a ``recombined spot shape'' criterion to disentangle false positives from true planets, and the importance of the observer's judgement for the reliability of this technique in its current implementation.

\begin{table}
  \begin{center}
    \begin{footnotesize}
    \begin{tabular}{ c | c | c | c }
      \hline
      \hline
      \multirow{2}{*}{Simulation} & Grid search & \multicolumn{2}{c}{Gradient descent} \\
      & S/N 100th value & Minimum S/N & Maximum S/N \\
      \hline      
      1 & 4.27 & 4.62 & 5.64 \\
      6 & 4.45 & 4.60 & 4.95 \\
      8 & 4.58 & 4.93 & 5.74 \\
      11 & 4.76 & 4.82 & 5.78 \\
      13 & 4.50 & 4.61 & 5.26 \\
      15 & 5.50 & 5.60 & 5.96 \\
      21 & 4.80 & 4.86 & 5.59 \\
      26 & 4.57 & 4.59 & 5.24 \\
      27 & 4.55 & 4.67 & 5.67 \\
      29 & 4.45 & 4.60 & 5.80 \\
      37 & 4.82 & 4.89 & 5.68 \\
      38 & 4.75 & 4.82 & 5.69 \\
      41 & 4.86 & 4.92 & 5.55 \\
      45 & 4.57 & 4.66 & 6.55 \\
      49 & 4.46 & 4.52 & 5.70 \\
      \hline
    \end{tabular}
    \caption{Values of the 100th best S/N ratio found by the grid search, and minimum/maximum values found by the reoptimization algorithm for each of the 15 simulations were no planet was injected.}
    \label{tab:no_planet_snr}
        \end{footnotesize}
  \end{center}
\end{table}

\subsection{Can we get rid of the astronomer?}

In its current implemention, the K-Stacker algorithm produces 100 images for each time serie analyzed, and requires the intervention of a skillful observer to check these recombined images for the presence of a planet. This step, while also used in classical high-contrast imaging, is tedious, and may introduce uncontrolled bias in the data reduction process. But for the time being, and as already discussed in Section~\ref{sec:false_positives}, this step is absolutely necessary to avoid large number of false positives.

\begin{figure}
  \begin{center}
    \includegraphics[width = \linewidth]{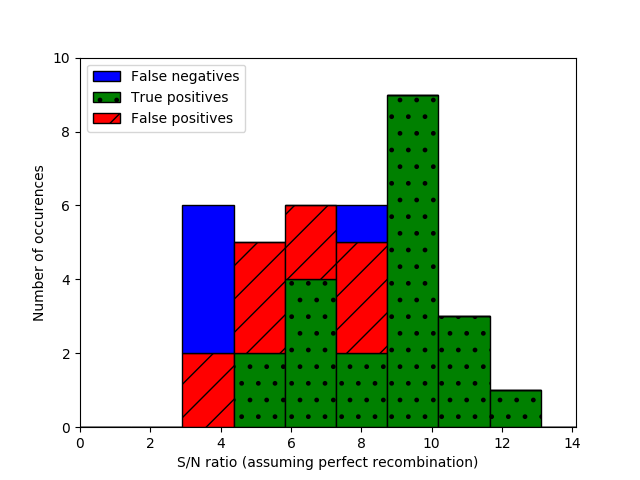}\\
    \caption{Resulting distribution of the planet candidates found and missed among the 33 simulations were a planet was effectively injected, when using only a S/N > 5 treshold to flag planet candidates.}
    \label{fig:histogram_snr_5}
  \end{center}
\end{figure}

\begin{figure}
  \begin{center}
    \includegraphics[width = \linewidth]{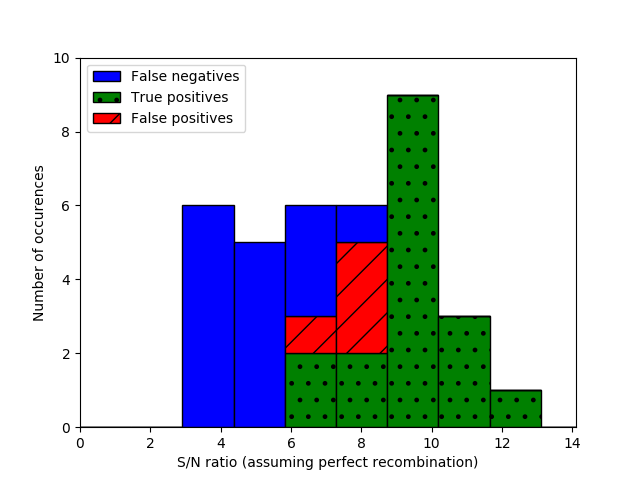}\\
    \caption{Same as Figure~\ref{fig:histogram_snr_5}, but for a S/N threshold of 7}
    \label{fig:histogram_snr_7}
  \end{center}
\end{figure}

In Figure~\ref{fig:histogram_snr_5} (resp. \ref{fig:histogram_snr_7}), we show what would have been the results of our blind test if we had used only a S/N threshold to flag planet candidates. For each simulation, the best solution found by K-Stacker is simply flagged as a planet candidate if the S/N is above 5 (resp. 7), without any intervention of an observer. As expected, the number of false positives (10, resp. 5 in total) is much higher, especially at low S/N. In our blind test, the observer was able to detect that these ``best solution'' found by K-Stacker were actually ``super-speckles'', and he eliminated them {(see also Fig. \ref{k-stacker images})}. For most of them, he was also able to find the planet candidates.

As discussed in Section~\ref{sec:improvements}, the shape of the recombined spot is important. To try to take that into account, we switched from our simple circular photometric aperture to a slightly more sophisticated gaussian weighted aperture. We found no real difference with our simple circular aperture, and this new method did not alleviate the need for the observer to carefully look at each image to assess the presence of the planet. A more complex algorithm, maybe one based on machine-learning trained to disentangle a planet PSF from recombined speckles, will be required in order to get rid of the observer's judgement. This would be a major improvement for K-Stacker, but left for future work. {Other astrophysical informations (spectral, polarimetry) will also be used to remove the false alarms.}

\subsection{Orbital parameters determination}

K-Stacker can also provides an estimate of the orbital parameters, as a by-product of the optimization algorithm. The precision with which these parameters are estimated depends on many different factors, such as the actual orbit on which the planet is moving, the total time spanned by the observing sequence, etc. It is clear, for example, that if the planet does not move significantly during the sequence, one should not expect to have reliable information about the orbital parameters. In such cases, K-Stacker is able to recenter the images and detect planets not reachable with other methods, but several very different sets of orbital parameters can lead to a good recombination. 

\begin{figure}
  \begin{center}
    \includegraphics[width = \linewidth]{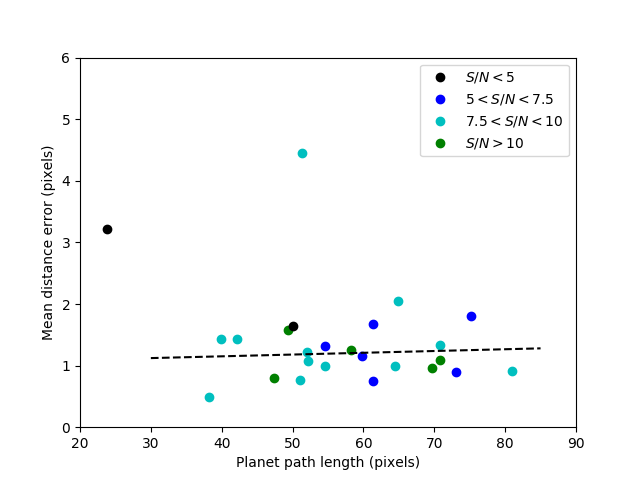}
    \caption{Mean distance between the true position of the planet and the position found by K-Stacker as a function of the total path length travelled by the planet, for the 25 simulations in which the planet has been detected.}
    \label{fig:precision}
  \end{center}
\end{figure}

In Figure~\ref{fig:precision}, we give, for each of these 25 simulations in which the planet was found, the mean distance between the true position of the planet in each image and the one predicted according to the set of parameters optimized by the K-Stacker algorithm, as a function of the total path travelled, and of the S/N ratio. The two points with a mean error of more than 3 pixels correspond to two simulations in which the planet is very close to the edge of the AO corrected area in certain images of the sequence, reducing the ability of the algorithm to properly constrain the orbit.

A linear fit on the other points reveal that the mean distance error is slightly increasing with the total path length travelled by the planet ($3\times{}10^{-3}$ pixel per pixel travelled by the planet).

When then planet travels along a large portion of its orbit in the sequence, we expect the orbit to be better constrained, and the algorithm to find a better fit to it. But at the same time, a fixed error on the orbital elements has a much bigger impact on the estimated positions if the planet moves along a large portion of its orbit. The result of Figure~\ref{fig:precision} shows that we are still dominated by the second effect, suggesting that the reoptimization stage could still be improved.

\begin{figure*}
  \begin{center}
    \includegraphics[width = \linewidth]{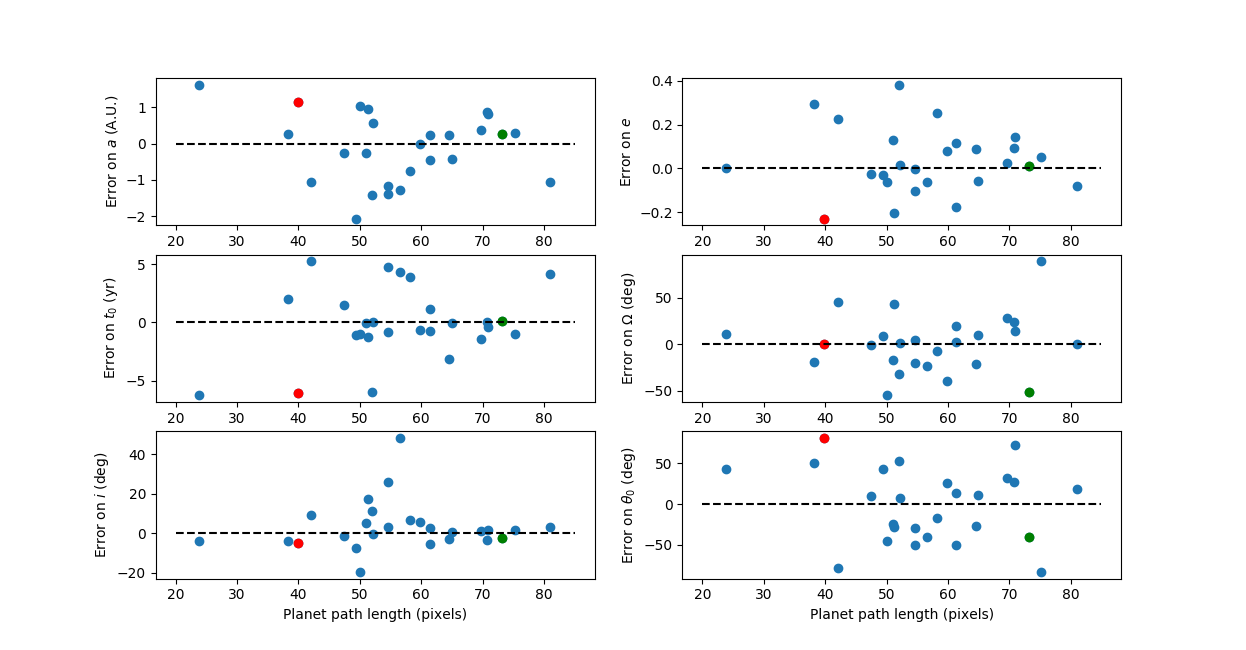}\\
    \includegraphics[width = 0.35\linewidth]{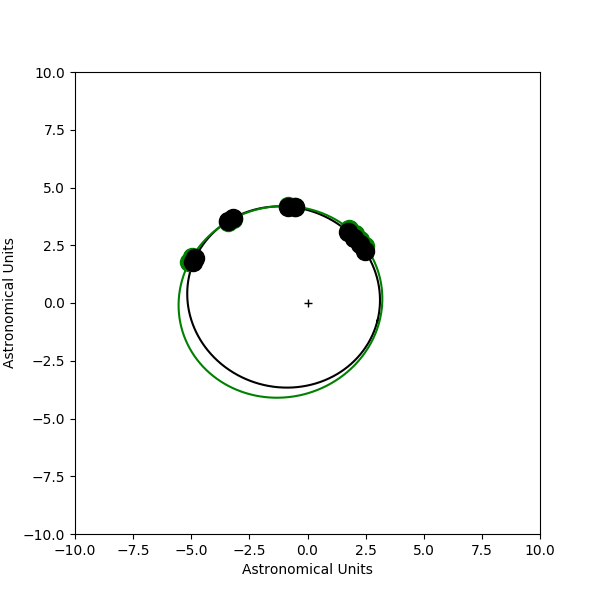}
    \includegraphics[width = 0.35\linewidth]{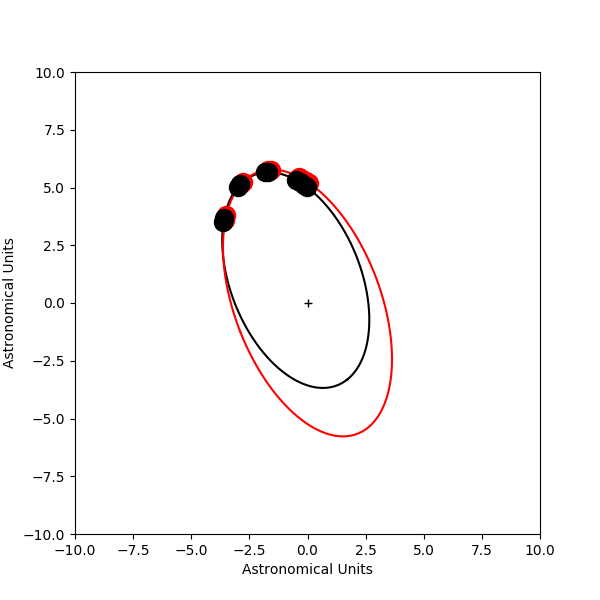}    
    \caption{The upper 6 panels show the errors made one the orbital parameters as a function of the total path travelled by the planet, for each of the simulation in which a planet has correctly been identified. In these panels, two cases, respectively corresponding to a good fit of the orbit (green dots) and to a bad fit (red dots) are highlighted. The corresponding orbits are shown in the two lower panels, with the real orbit of the planet in black, and the orbit found by K-Stakcer in green (good fit case), and red (bad fit case). For each orbit, the position of the planet in all the 10 images is also represented.}
    \label{fig:param_errors}
  \end{center}
\end{figure*}

In Figure~\ref{fig:param_errors}, we show the error on each of the 6 parameters individually, as a function of the total path length travelled by the planet. On some parameters (e.g. $a$, $i$, $t_0$), there seems to be a clear advantage of having longer displacements of the planet to get better estimates. On some others (e. g. $e$, $\Omega$, $\theta_0$), it is not as clear, though. These parameters, which are not better fitted for longer travelled path, may be responsible for the overall increasing slope of Figure~\ref{fig:precision}.

\subsection{Possible improvements of the algorithm}
\label{sec:improvements}

We find that the current version of the K-Stacker algorithm can achieve a $50~\%$ detection rate on targets with a total \hbox{$\mathrm{S/N}\simeq{}5$}.
Whereas this result already shows that it is possible to use K-Stacker to find very faint planets (S/N below 2 in individual frames for a serie of 10 observations), we strongly believe that there is still room for improvement in the algorithm.

One sure way to improve the results of the K-Stacker algorithm would be to optimize the search-grid. For example, a larger steps in $\theta_0$ could be used for smaller values of $a$. In our version of the algorithm, we are exploring the same range of values for $t_0$ no matter what the actual value of $a$ is. But it is well known that the semi major-axis and the orbital period are directly related, and this could be used to reduce the range explored for smaller values of $a$.

{In fact, an optimal grid may be constructed for the K-Stacker algorithm by looking for a set of orbit which diverge by at least one FWHM of the instrument PSF in at least one image of the sequence. This could be done numerically, or maybe even analytically. This would ensure that the minimum possible number of orbits are used, and hence would help in reducing the total computing time, and the false alarm rate.}

Also, one area which has yet to be studied in details is the definition of the function to be optimized usually refered to as the ``gain function'' in optimization problems. As pointed out in Sect.~\ref{sec:false_positives}, when the algorithm is trying to recombine speckles to create a ``super-speckle'', the resulting pattern is ``blurred'' and does not ressemble a real PSF. The most efficient gain function might be one that takes into account the S/N but also the shape of the recombined spot. {When searching for planets close to the star, small sample statistics should be taken into account to compute the S/N \citep{Mawet2014}.}

{Along this line of thought, we also tried to use a noise-weighted averaging of the different images (close to what is suggested in \citeauthor{Bottom2017}, \citeyear{Bottom2017}), to take into account the fact that in our series, because of the motion of the planet with respect to the central star, some exposures might be better than others (e.g. when the planet is far away from the star). This modification did not drastically changed our results. The same blind test lead to 25 planets recovered out of 34 (instead of the 24 in the version presented here), with no false negative instead of one found here. A more sophisticated algorithm will be necessary to really improve the results.}

\begin{figure}
  \begin{center}
    \includegraphics[width=0.9\linewidth]{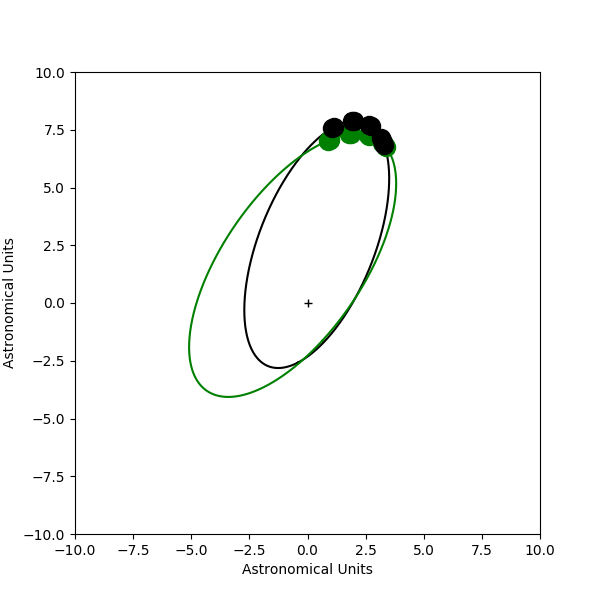}
    \includegraphics[width=0.9\linewidth]{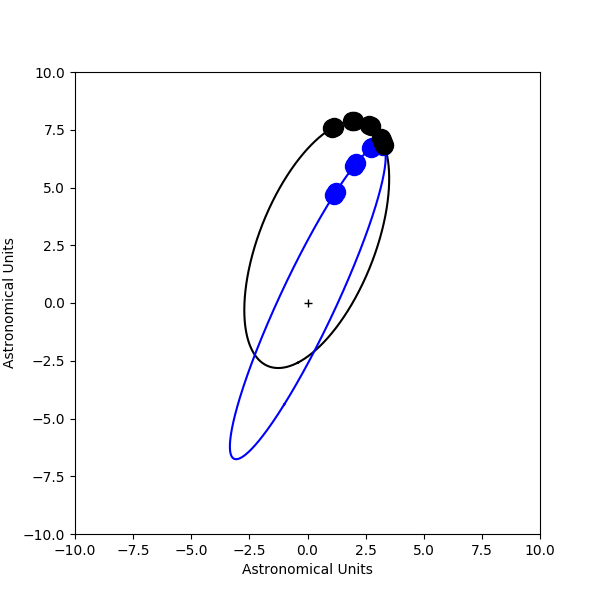}
    \caption{Illustration of an unexpected behavior of the K-Stacker algorithm when using a noise-weighted averaged combination of images. The top panel shows the best orbit found by the classical combination algorithm, and the bottom panel shows the best orbit found when using the noise-weighted averaged combination, for the same simulation. In this case, the planet has a S/N $\simeq{} 1.9$ in the first 4 images, and $\simeq{} 1.2$ in the last 6 images. The real orbit of the planet is in black.}
    \label{fig:orbit_combination}
\end{center}
\end{figure}

{We also noted an unexpected behavior of the algorithm when using the noise-weighted average. This behavior is examplified in one particular simulation, in which the planet was found by the algorithm, but the orbit was not properly recovered (see Figure~\ref{fig:orbit_combination}). An in-depth investigation revealed that this was due to some sort of interaction between the way the planets are injected in our images, and the noise-weighted averaging. We recall that in our images, the planet is injected at constant flux, rather than constant S/N, to reflect the reality of a constant exposure time. This means that from one image to the other, the actual S/N may vary. In this particular case, the planet was injected at a higher S/N in the first few images of the sequence than in the others (S/N$\simeq{}$1.9 vs 1.2). Because of the noise structure in high-contrast images, the noise-averaged combination also has a natural tendancy to favor images in which the computed position is far from the central star. The combination of these two factors made for a situation in which a highly eccentric orbit fitting correctly the first part of the real orbit and moving rapidly towards the central star in the last part (as in Figure~\ref{fig:orbit_combination}, bottom panel) yields a good total S/N. The classical combination algorithm, which gives the same importance to each image of the sequence, is less subject to this effect.}

Finally, it has to be noted that the K-Stacking method does not necessarily implies the use of a brute-force optimization algorithm. Any type of optimization algorithm could be used to optimize the S/N function: simulated annealing, genetic algorithms, amoeba, etc. However, we believe that the brute-force method is one of the most appropriate for K-Stacker. Despite not being known for its efficiency, this method has a very interesting property: the sets of orbital parameters on which the S/N has to be calculated are known in advance and never change (these are the points of the search grid). This means that adding a new image to a set of $n$  observations already processed by K-Stacker do not take much more computation time if the ``signal'' and ``noise'' terms computed in each image and for each point of the search grid are systematically stored. In this case, these terms can be combined to generate the S/N values using the equation $\mathrm{S/N}=\sum_k S_k / \sqrt{\sum_k \sigma{}_k^2}$. Then to add a $n+1$th image to a set of $n$ images, one can just compute the values of $S_{n+1}$ and $\sigma_{n+1}$, and recompute the S/N. The new image can be processed separately and it is not necessary to reprocess the entire set of $n+1$ images. On the opposite, when using Monte-Carlo methods, in which the optimization path is generated dynamically by the algorithm based on previous values found, adding a new image to a set of already processed observations does require a complete re-processing of the whole set, which can prove extremely time-consuming.

{On the other hand, a Monte-Carlo type algorithm would explore more thoroughly the local search space around the best orbits, helping in determining error bars on the orbital parameters. The best way to proceed might be a combination of both a grid search to reduce the search space, and an MCMC (Monte-Carlo Markov Chain) based algorithm.}

\section{Conclusion: what can we use K-Stacker for?}
\label{sec:conclusion}

In this study, we have shown how the K-Stacking method could be used to combine multiple high-contrast images obtained during different nights to detect exoplanets. We simulated SPHERE/IRDIS coronagraphic images taken over several months with planets at low S/N level (below the detection limit) in individual frames, and used the K-Stacker algorithm to combine these images and detect the planets. We find that when the total number $n$ of available images is large enough to get $\sqrt{n}\times{}\text{S/N}_{ind} \ge{} 7$ (where $\text{S/N}_{ind}$ is the S/N levels in individual frames), the K-Stacker algorithm is able to detect the planet with a very high level of reliability ($> 90\%$). For cases where $5\le \sqrt{n}\times{}\text{S/N}_{ind} \le 6$, we find a recovery rate of about 50 \%. 

The detection rate can probably be improved by refining and optimizing the search grid. This will be done in a future work, using real data. {This first study was limited to simulated images in order to have complete control over all the important parameters (star magnitude, turbulence conditions, XAO performances, etc.)}. It has shown that K-Stacker could achieve good recovery rates (that is: simillar to what usual high-contrast imaging techiques are expected to provide) on simulated data, in a decent amount of time. Future studies on this subject should be done using real data, possibly with flase planets injected. 

In its current implementation, K-Stacker still heavily relies on the ability of the observer to identifiy false positives in the recombined images. Whereas our blind test has shown that this could be done reliably (only one false positive, which was identified as an uncertain planet candidate by the observer, among our 49 simulations), it is clear that this has to be tested using real data.

In its current state, though, the K-Stacker technique could already prove useful to detect very faint planets. The ADI technique instrinsically limits the useful observing time which can be spent each night on any single star every night (to about 1 - 2 hr). Adding to this that high-contrast imaging instruments require the best observing conditions, that these instruments are sharing telescope time with others, and that stars are not visible year-round in the night sky, and accumulating about 10 hr of ADI observation on a single target in a few months (to limit orbital motion) can prove exessively complicated. Using the K-Stacker would drastically improve the chances of doing so, and reduce the complexity of observing schedules.

Finally, we also want to point out that this method could also be used as part of a new scheme of observation, in which exposures would not be made sequentially in one night, but would be spread over multiple nights, to better contsraint the orbit. For the same contrast in the final image, the K-Satcker technique would also yield an estimate of the orbital parameters, which cannot be obtained with a single night observation.

\begin{acknowledgements}
  We thank the anonymous referee for his/her careful reading of the manuscript, and his/her helpful suggestions and comments. This article was significantly improved after his/her review.  \\
  This research has been funded by the PNP/ASHRA/INSU. This paper has used the cluster facilities of LAM operated by the CeSAM data center.
\end{acknowledgements}

\bibliographystyle{aa} 
\bibliography{biblio} 

\appendix
\section{Example of solutions found by K-Stacker}
\label{app:solutions}

\begin{figure*}
  \subfigure[]{\label{sub1} \includegraphics[width=9cm]{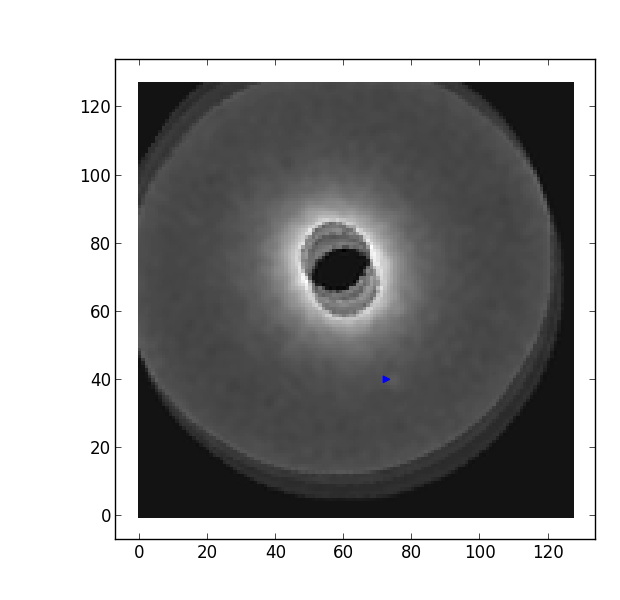}}
  \subfigure[]{\label{sub2} \includegraphics[width=9cm]{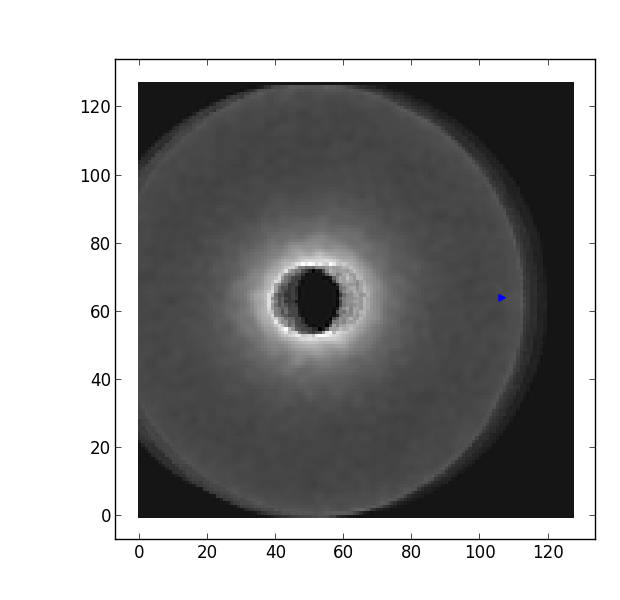}}\\
  \subfigure[]{\label{sub3} \includegraphics[width=9cm]{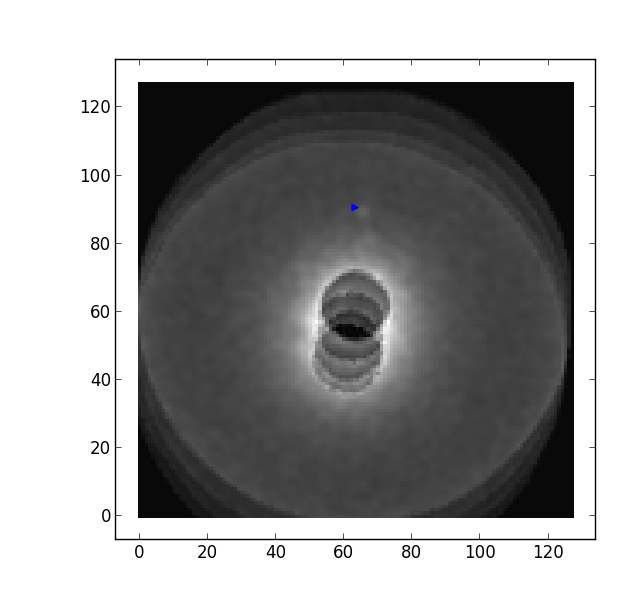}}  
  \subfigure[]{\label{sub4} \includegraphics[width=9cm]{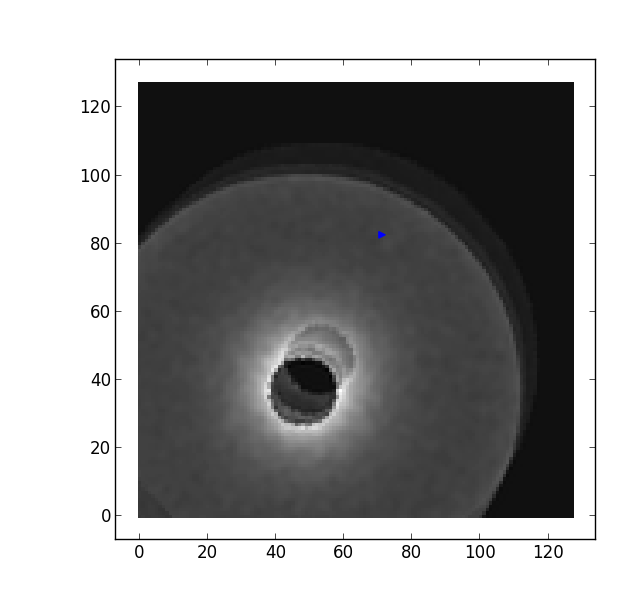}}
\end{figure*}
\begin{figure*}
  \subfigure[]{\label{sub5} \includegraphics[width=9cm]{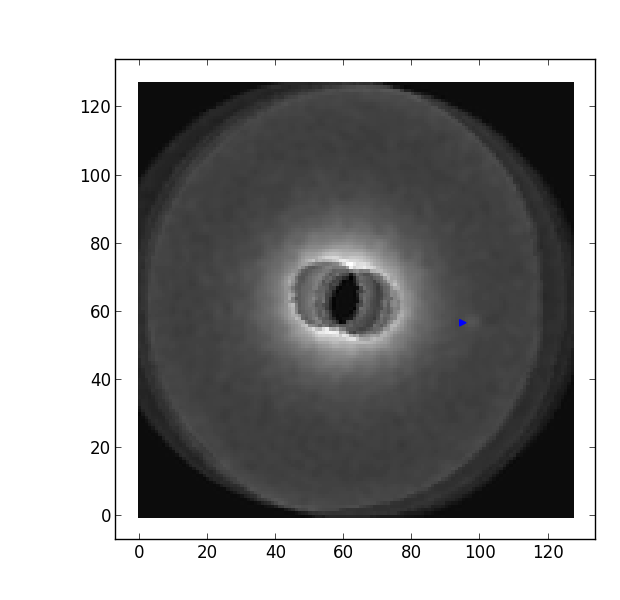}}
  \subfigure[]{\label{sub6} \includegraphics[width=9cm]{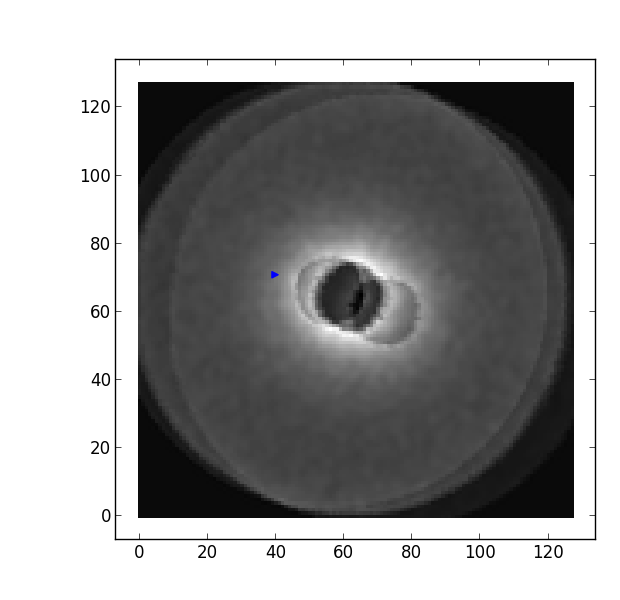}}        \\
  \subfigure[]{\label{sub7} \includegraphics[width=9cm]{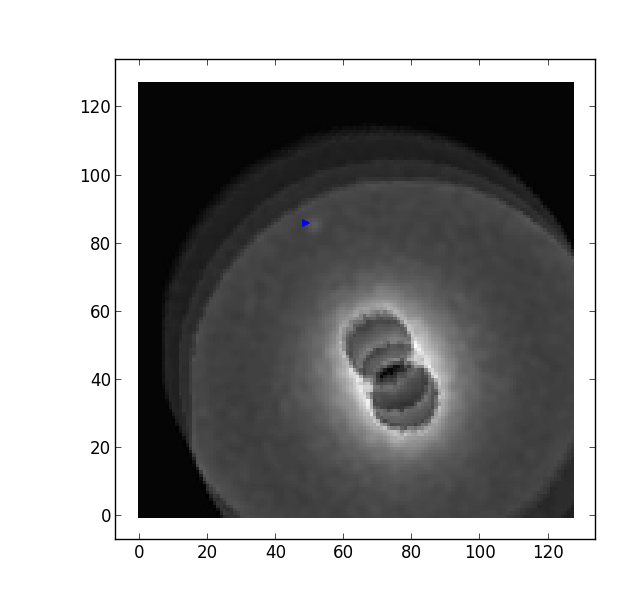}}
  
  \caption{A set of best recombined images as found by K-Stacker, for different reoptimized S/N values, and for cases where the algorithm found a true/false planet candidate. In each panel, the blue arrow show the position of the plantet candidate found by K-Stacker. Panel \ref{sub1} (resp. \ref{sub2}):  S/N = 5.7 with a correct (resp. false) planet candidate; Panel \ref{sub3} (resp. \ref{sub4}):  S/N = 6.0 with a correct (resp. false) planet candidate; Panel \ref{sub5} (resp. \ref{sub6}):  S/N = 6.5 with a correct (resp. false) planet candidate;  Panel \ref{sub7}: S/N = 8.3, with a true planet candidate. Note that the observer has been able to correctly identify false alarm ($S/N>6$ without planets injected in simulations: \ref{sub2}, \ref{sub4}, and \ref{sub5}) using the shape of the spot, without any a priori information on the planet injection.}
  \label{k-stacker images}
\end{figure*}

\end{document}